\documentclass{LMCS}

\usepackage{enumerate}
\usepackage{amsfonts, amsmath, amstext, amssymb}
\newtheorem{lemma}[thm]{Lemma}
\newtheorem{defn}[thm]{Definition}
\newtheorem{Question}{Question}

\newcommand{\Q}{\mathbb{Q}}
\newcommand{\R}{\mathbb{R}}
\newcommand{\N}{\mathbb{N}}
\newcommand{\Z}{\mathbb{Z}}
\newcommand{\Qbar}{\overline{\mathbb{Q}}}
\newcommand{\A}{\mathbb{A}}

\newcommand{\C}{\mathbb{C}}
\newcommand{\D}{\mathcal{D}}

\newcommand{\M}{\mathcal{M}}
\newcommand{\bH}{\mathbb{H}}
\renewcommand{\P}{\mathcal{P}}

\def\res{\!\!\upharpoonright\!}

\def\phi{\varphi}

\newcommand{\la}{\langle}
\newcommand{\ra}{\rangle}

\newcommand{\ep}{\epsilon}
\newcommand{\fC}{\mathfrak{C}}

\newcommand{\Gal}[2]{\text{Gal}(#1/#2)}

\newcommand{\xbar}{\overline{x}}
\newcommand{\sigmabar}{\overline{\sigma}}
\newcommand{\Sbar}{\overline{S}}

\newcommand{\sqrtu}{\sqrt{u}}

\newcommand{\avec}{\vec{a}}
\newcommand{\bvec}{\vec{b}}
\newcommand{\pvec}{\vec{p}}
\newcommand{\tvec}{\vec{t}}
\newcommand{\Tvec}{\vec{T}}
\newcommand{\xvec}{\vec{x}}

\newcommand{\yvec}{\vec{y}}
\newcommand{\Yvec}{\vec{Y}}
\newcommand{\zvec}{\vec{z}}
\newcommand{\set}[2]{\ensuremath{ \{ #1 : #2 \} }}

\renewcommand{\deg}[1]{\text{deg}(#1)}
\newcommand{\dom}[1]{\text{dom}(#1)}

\theoremstyle{plain}

\def\doi{7 (2:15) 2011}
\lmcsheading%
{\doi}
{1--20}
{}
{}
{Oct.~30, 2010}
{May\phantom.~24, 2011}
{}
\begin{document}

\title[Noncomputable Functions in the Blum-Shub-Smale Model]{Noncomputable Functions\\in the Blum-Shub-Smale Model\rsuper*}

\author[W.~Calvert]{Wesley Calvert\rsuper a}
\address{{\lsuper a}Department of Mathematics\\
Mail Code 4408\\
Southern Illinois University\\
1245 Lincoln Drive\\
Carbondale, Illinois 62901}
\email{wcalvert@siu.edu}
\urladdr{http://www.math.siu.edu/calvert}
\thanks{{\lsuper a} Partially supported by Grant \#13397 from the Templeton Foundation.}

\author[K.~Kramer]{Ken Kramer\rsuper b}
\address{{\lsuper b}Department of Mathematics\\
Queens College -- C.U.N.Y.\\
65-30 Kissena Blvd.\\
Flushing, New York 11367 U.S.A.;
Ph.D.\ Program in Mathematics\\
C.U.N.Y.\ Graduate Center\\
365 Fifth Avenue\\
New York, New York 10016 U.S.A.}
\email{kkramer@qc.cuny.edu}
\thanks{{\lsuper b}Partially supported by NSF grant \# DMS-0739346.}

\author[R.~Miller]{Russell Miller\rsuper c}
\address{{\lsuper c}Department of Mathematics\\
Queens College -- C.U.N.Y.\\
65-30 Kissena Blvd.\\
Flushing, New York 11367 U.S.A.;
Ph.D.\ Programs in Mathematics \& Computer Science\\
C.U.N.Y.\ Graduate Center\\
365 Fifth Avenue\\
New York, New York 10016 U.S.A.}
\email{Russell.Miller@qc.cuny.edu}
\urladdr{http://qc.edu/\textasciitilde rmiller}
\thanks{{\lsuper c}Partially supported by Grant \#13397 from the Templeton
Foundation, by NSF grant \# DMS-1001306,
by grants numbered 61467-00 39, 62632-00 40, and 63286-00 41 from
The City University of New York PSC-CUNY Research Award Program,
and by Queens College Research Enhancement Program award \# 90927-08 08.}



\titlecomment{{\lsuper*}Portions of this article describe results which
appeared as \cite{CKMCCA10} in the conference proceedings volume
of the meeting \emph{Computability and Complexity in Analysis}
in Zhenjiang, China, 21-25 June 2010, and other results
which were presented at the meeting \emph{Logical Approaches
to Barriers in Computing and Complexity} in Greifswald, Germany, 17-20 February 2010.}


\begin{abstract}
\noindent Working in the Blum-Shub-Smale model of computation
on the real numbers, we answer several questions of Meer
and Ziegler.  First, we show that, for each natural number $d$,
an oracle for the set of algebraic
real numbers of degree at most $d$ is insufficient to allow
an oracle BSS-machine to decide membership in the set of algebraic
numbers of degree $d+1$.  We add a number of further
results on relative computability of these sets and their unions.
Then we show that the halting problem for BSS-computation
is not decidable below any countable oracle set,
and give a more specific condition, related to the
cardinalities of the sets, necessary for relative BSS-computability.
Most of our results involve the technique of using
as input a tuple of real numbers which is algebraically
independent over both the parameters and the oracle
of the machine.
\end{abstract}

\maketitle


\section{Introduction}
\label{sec:intro}

Blum, Shub, and Smale introduced in \cite{BSS} a notion of computation
with full-precision real arithmetic, in which the ordered field
operations are axiomatically computable, and the computable functions
are closed under the usual operations.  A complete account of
this model is given in \cite{BCSS}.  A program for such a machine consists
of a finite set of instructions as described there, and the instructions
are allowed to contain finitely many real parameters, since a single real
number is viewed as a finite object.  The program can
add, multiply, subtract, or divide real numbers in its cells,
can copy cell, can overwrite the contents of a cell with $0$, and can
use the relations $=$ and $<$ to compare the contents of two cells,
forking according to whether the contents of those cells
satisfy that relation.  For our purposes, it will
be convenient to assume that the forking instructions in the program
compare the real number in a single given cell to $0$, under either
$=$ or $<$ or $>$.  Such a machine has equivalent computing power
to machines which can compare the contents of two different cells
to each other.

Of course, the BSS model is not the only concept of computation on $\R$,
nor should it be considered the dominant model.  It corresponds
to a view of the real numbers as a fixed structure, perhaps given
axiomatically --  defined, for instance, as the unique complete Archimedean
ordered field, with field operations vouchsafed unto us mathematicians;
as opposed to a view of real numbers as objects defined by
Cauchy sequences or by Dedekind cuts in the rational numbers $\Q$,
with operations derived from the analogous operations on $\Q$.
There is no obvious method of implementing
BSS machines by means of digital computers.  More typically, as in
\cite{BSS,BCSS}, one envisions BSS machines as a model for numerical
computation in which features of approximation, rounding, and error
analysis are treated as a separate posterior analysis.
This failure invites a contrast with computable analysis,
which treats real numbers as quantities
approximated by rational numbers and is intended to reflect the capabilities of
digital computers.   However, the BSS model is of interest
both for the analogy between it and the Turing model, which can be seen as
BSS computation on the ring $\Z/(2)$, and because it reflects
the intuitions of many mathematicians -- dating back to the nineteenth
century, and mostly outside of computer science -- about
the notion of algorithmic computation on $\R$.
Some useful further discussion of these questions
appears in \cite{W99}.

This paper will consider sets of algebraic real numbers, and
other sets of tuples from $\R$, as oracles for BSS machines,
and will examine
the relative difficulty of deciding membership in such sets under the
BSS model of computation.  Several sections compare various sets
of algebraic numbers under BSS-oracle computation,
using these sets to demonstrate that there exists a rich structure
of BSS-semidecidable degrees under BSS reducibility.
Later sections consider questions
about cardinality:  to what extent the complexity of a subset of $\R$ (or $\C$)
allows us to draw conclusions about its cardinality.  The previous paper \cite{MZ08}
by Meer and Ziegler focused attention on these issues, and here
we answer several of the questions raised there.  Our method
adapts a known technique from BSS computability, and should
be comprehensible to casual readers as well as to logicians
and computer scientists.  It requires significant use of algebraic
properties of the real numbers, in addition to computability,
reinforcing the general perception of the BSS model as an
essentially algebraic approach to computation on $\R$,
treating real numbers as indivisible finite items.
In contrast, the use of computable analysis normally results
in a more analytic approach to computation on $\R$.
We the present authors comprise a number theorist
and two computable model theorists with experience in algorithms
on (countable) Turing-computable fields, and thus we
are more familiar with the algebraic side.

Our notation generally follows that of \cite{MZ08}.
The set of all finite tuples of real numbers is denoted
$\R^\infty$; the inputs and outputs of BSS machines on $\R$
all lie in this set, and the collective content of the cells of a BSS
machine at a given stage in a computation may also be regarded
as an element of $\R^\infty$.  We use $\A$ to denote the set of all
real numbers which are algebraic over the subfield $\Q$
of rational numbers.  $\A$ is partitioned into subsets $\A_{=d}$,
for each natural number $d$, so that $\A_{=d}$ contains those
algebraic real numbers of degree exactly $d$ over $\Q$.
(Recall that the \emph{degree} of $x$ over $\Q$ is the
vector space dimension over $\Q$ of the field $\Q(x)$
generated by $x$; equivalently, it is the degree of the
minimal polynomial of $x$ in $\Q[X]$.)  We also write
$\A_{\leq d}=\cup_{c\leq d}\A_{=c}$, the set of algebraic real
numbers of degree $\leq d$.  (In \cite{MZ08}, this set was
called $\A_d$; our notation is intended to distinguish
$\A_{=d}$ from $\A_{\leq d}$.)  By the definition of degree,
$\A_{=0}$ is empty, and $\A_{=1}$ contains exactly the rational
numbers themselves.  We mention \cite{vdW70} as an excellent
source for these and other algebraic preliminaries,
and \cite{FJ86} for more advanced questions about algorithms
on fields.

The following lemma is well known, and clear by induction on stages.
It reflects the fact that the four field operations are the only
operations which a BSS machine is able to perform.
\begin{lemma}
\label{lemma:fieldboundary}
If $M$ is a BSS machine using only the real parameters $\zvec$
in its program, then at every stage of the run of $M$ on any input
$\xvec$, the content of every cell lies in the field $\Q(\zvec,\xvec)$.
\qed
\end{lemma}

It is immediate from this lemma that the set $\A$
cannot be the image of $\N$ under any BSS-computable function,
as it is not contained within any finitely generated field. (We will
tend to use $\N$ to denote the subset of $\R$
consisting of the nonnegative integers, as here,
whereas $\omega$ will denote the same set when not sitting inside of $\R$.)
We say that $\A$ is not \emph{BSS-countable}.  On the other hand,
$\A$ does satisfy the definition of \emph{BSS semidecidability},
which is the best analogue of Turing-computable
enumerability and has been studied more closely in the literature.
\begin{defn}
\label{defn:BSSsemidecidable}
A set $S\subseteq\R^\infty$ is \emph{BSS-semidecidable} if there
exists a (partial) BSS-computable function with domain $S$,
and \emph{BSS-countable} if there exists a partial BSS-computable
function mapping $\N$ onto $S$.  A set $S$ is \emph{BSS-decidable}
if its characteristic function $\chi_S$ is BSS-computable.
\end{defn}
It is immediate that $S$ is BSS-decidable if and only if both
$S$ and $(\R^\infty -S)$ are BSS-semidecidable.  This justifies
the analogy between BSS-semidecidability in $\R^\infty$ and
computable enumerability in $\omega$, and also dictates
the use of the prefix ``semi.''  The term \emph{BSS-countable},
on the other hand, suggests that the set can be listed out,
element by element, by a BSS machine, which is precisely the
content of the definition above.  (The related term \emph{BSS-enumerable}
has been used by other authors to denote the image
of $\R^\infty$ under a BSS-computable partial function.)
In the context of Turing computation,
computable enumerability and semidecidability are equivalent, but
in the BSS context, the set $\A$ distinguishes
the two notions, being BSS-semidecidable but not BSS-countable.
(On the other hand, every BSS-countable set
is readily seen to be BSS-semidecidable.)
The semidecision procedure for $\A$ is well-known:
take any input $x$, and go through all nonzero polynomials
$p(X)\in\Q[X]$, computing $p(x)$ for each.  If ever $p(x)=0$,
the machine halts.  The ability to go through the polynomials
in $\Q[X]$ follows from the BSS-countability of $\Q[X]$,
which in turn follows from the BSS-countability of $\Q$.
(A similar result applies to the set of algebraically dependent
tuples in $\R^\infty$; see for instance \cite{KZ08}.)

The two questions which gave rise to this paper were posed by Meer and Ziegler in
\cite{MZ08}.  Both use the notion of a \emph{BSS reduction}, analogous to Turing reductions.
An \emph{oracle BSS machine} is essentially a BSS machine with the additional
ability to take any finite tuple (which it has already assembled on the cells of its tape),
ask an oracle set $A$ whether that tuple lies in $A$, and fork
according to whether the answer is positive or negative.
The oracle $A$ should be a subset of $\R^\infty$, of course,
and we will write $M^A$ to represent an oracle BSS program
(or machine) equipped with an oracle set $A$.  More
precisely, \begin{defn} An oracle BSS machine using an oracle set $B
  \subseteq \R^\infty$ is a BSS machine with an additional type of
  node called an oracle node.  This node branches according as $x \in
  B$, where $x$ is some previously computed element.\end{defn}
(This is exactly the definition given in \cite{MZ08}, and is
equivalent to any reasonable formalization of the implicit definition
given in Problem 10.2 of \cite{BSS}.)\ \ 
Oracle BSS programs can be enumerated (by tuples from
$\R^\infty$) in much the same manner as regular BSS programs.
If $B\subseteq\R^\infty$ and the characteristic function $\chi_B$
can be computed by an oracle BSS machine $M^A$ with oracle $A$,
then we write $B\leq_{BSS} A$, and say that $B$ is \emph{BSS-reducible}
to $A$, calling $M$ the \emph{BSS reduction} of $B$ to $A$.
Should $B\leq_{BSS} A$ and also $A\leq_{BSS}B$,
we write $A\equiv_{BSS}B$ and call the two sets
\emph{BSS-equivalent}.  All this is exactly analogous to oracle
Turing computation on subsets of $\omega$.  The first question regards
the connection of this reducibility with algebra.

\begin{Question}\label{mzq} Let $\mathbb{A}_{\leq d}$ be the set of algebraic numbers
with degree (over $\mathbb{Q}$) at most $d$.  Then is it true that
\[\mathbb{A}_{\leq 0} <_{BSS} \mathbb{A}_{\leq 1} <_{BSS}
\cdots \mathbb{A}_{\leq n} <_{BSS} \cdots?\]\end{Question}

That $\A_{\leq d-1}\leq_{BSS}\A_{\leq d}$ is immediate for all $d$;
see Lemma \ref{lemma:SinT} below.  The focus
of the question is on the lack of any reduction in the opposite direction.

Meer and Ziegler credit the second question to an anonymous referee of \cite{MZ08}.

\begin{Question}\label{mzqc} Let $\mathbb{A}$ be the set of algebraic numbers
in $\R$, i.e.\ those which are roots of a nonzero polynomial in $\Q[X]$.
Also, let $\bH$ be the Halting Problem for BSS computation on $\R$,
as described in \cite{MZ08} (Actually, a passing implicit reference is
made to this set in \cite[\S8]{BSS}, in the guise of the halting set
of a universal machine).  Is it true that $\bH \not\leq_{BSS} \A$?
And more generally, could any countable subset of $\R^\infty$
contain enough information to decide $\bH$?
\end{Question}

That $\A\leq_{BSS}\bH$ is immediate.
Let $P$ be the BSS program which, on input $x\in\R$,
plugs $x$ successively into each nonzero polynomial $p(X)$
in (the BSS-countable set) $\Q[X]$ and halts if ever $p(x)=0$.
Then $x\in\A$ iff the program $P$ halts on input $x$.
(Similarly, every BSS-semidecidable set is BSS-decidable in $\bH$,
and indeed $1$-reducible to $\bH$ in the BSS model.)  Again,
the question focusses on the lack of any reduction in the opposite direction.

Section \ref{sec:basic} gives the basic technical lemma
used in this paper to address such questions, and Section
\ref{sec:general} applies it to give a positive answer to
Question \ref{mzq}.  To aid the reader's comprehension,
Section \ref{sec:quadratic} describes the solution to Question
\ref{mzq} in the specific case $\A_{\leq 2} <_{BSS} \A_{\leq 1}$.
Section \ref{sec:hopeful} considers other possible
reductions among the sets $\A_{=d}$ for different values
of $d$, and among unions of these sets.  As a corollary,
we define a new embedding of the partial order $(\P(\omega),\subseteq)$
into the partial order of the BSS-semidecidable degrees 
under BSS reducibility.  Such embeddings, including the
similar one already derived in \cite{MZ08}, reveal a high level
of complexity within the latter structure.  In Section
\ref{sec:oracle}, we answer Question \ref{mzqc}
by showing that no countable subset of
$\R^\infty$ contains enough information to decide the
halting problem in the BSS model.  We also prove
there a theorem relating BSS degrees to cardinality,
showing that for infinite subsets $S\subseteq\R$ and
$C\subseteq\R^\infty$, if $S\leq_{BSS} C$, then the local cardinality
(in a technical sense defined in that section) of $S$
cannot be greater than the (global, i.e.\ usual) cardinality of $C$.
Finally, in Section \ref{sec:complex}, we offer analogies
of these observations for BSS computation on the
complex numbers, where the situation is far less messy.

\section{BSS-Computable Functions At Transcendentals}
\label{sec:basic}

Here we introduce our basic method for showing
that various functions on the real numbers fail to be
BSS-computable.  In Sections \ref{sec:quadratic} and \ref{sec:general},
this method will be extended to give answers
about BSS-computability below certain oracles.
However, even the non-relativized version yields
straightforward proofs of several well-known
results about BSS-decidable sets, as we will
see shortly after describing the method.

In many respects, our method is equivalent to
the method, used by many others, of considering
BSS computations as paths through a finite-branching
tree of height $\omega$, branching
whenever there is a forking instruction in the program.
However, we think that the intuition for our method
can be more readily explained to a mathematician
unfamiliar with computability theory.
Our main lemma says that near any transcendental input
in its domain, the values in a BSS-machine computation
must be defined by rational functions of the input.
Where previous proofs usually made arguments about
countable sets of terminal nodes in the tree
of possible computations, we simply use
the transcendence of this element.
\begin{lemma}
\label{lemma:epnbhd}
Let $M$ be a BSS-machine, and $\zvec$ the finite tuple
of real parameters mentioned in the program for $M$.
Suppose that $\yvec\in\R^{m+1}$ is a tuple of real numbers
algebraically independent over the field $Q=\Q(\zvec)$,
such that $M$ converges on input $\yvec$.  Then there
exists $\ep>0$ and rational functions $f_0,\ldots,f_n\in Q(\Yvec)$,
(that is, rational functions of the variables $\Yvec$
with coefficients from $Q$) such that for all
$\xvec\in \R^{m+1}$ with $|\xvec-\yvec|<\ep$,
$M$ also converges on input $\xvec$ with output
$\la f_0(\xvec),\ldots,f_n(\xvec)\ra\in\R^{n+1}$.
\end{lemma}
\begin{Proof}
The intuition is that by choosing $\xvec$ sufficiently close to $\yvec$,
we can ensure that the computation on $\xvec$ branches in
exactly the same way as the computation on $\yvec$,
at each of the (finitely many) branch points in the
computation on $\yvec$.  More formally,
say that the run of $M$ on input $\yvec$ halts at stage $t$,
and that at each stage $s\leq t$, the non-zero
cells contain the reals $\la f_{0,s}(\yvec),\ldots, f_{n_s,s}(\yvec)\ra$.
Lemma \ref{lemma:fieldboundary} shows that all
$f_{i,s}(\yvec)$ lie in the field $Q(\yvec)$, so each $f_{i,s}$
may be viewed as a rational function of $\yvec$ with
coefficients in $Q$.  Indeed, each rational function
$f_{i,s}$ is uniquely determined in $Q(\Yvec)$,
since $\yvec$ is algebraically independent
over $Q$.

Let $F$ be the finite set
$\set{f_{i,s}(\Yvec)}{s\leq t~\&~i\leq n_s~\&~f_{i,s}\notin Q}$
of nonconstant rational functions used in the computation.
The union $U$ of all preimages $f_{i,s}^{-1}(0)$ with
$f_{i,s}\in F$ is closed in $\R^{m+1}$,
and by algebraic independence, $\yvec$ does not lie in $U$,
so there exists an $\ep >0$ such that the $\ep$-ball
$B_\ep(\yvec)=\set{\xvec\in\R^{m+1}}{|\xvec-\yvec|<\ep}$,
does not intersect $U$,
and is contained within the domain of each $f_{i,s}\in F$.
Indeed,
for all $f_{i,s}\in F$ and all
$\xvec\in B_\ep(\yvec)$, $f_{i,s}(\xvec)$ and
$f_{i,s}(\yvec)$ must have the same sign,
since $B_{\ep}(\yvec)$ is a path-connected set.

Now fix any $\xvec\in B_\ep (\yvec)$.  We claim
that in the run of $M$ on input $\xvec$, at each stage $s\leq t$,
the cells will contain precisely $\la f_{0,s}(\xvec),\ldots,f_{n_s,s}(\xvec)\ra$
and the machine will be in the same state in which it was
at stage $s$ on input $\yvec$.  This is clear for
stage $0$, and we continue by induction, going
from each stage $s<t$ to stage $s+1$.  If the machine
executed a copy instruction or a field operation in this
step, then the result is clear, by inductive hypothesis.
Otherwise, the machine executed a fork instruction,
comparing some $f_{i,s}(\xvec)$ with $0$.  But we saw above
that $f_{i,s}(\xvec)$ and $f_{i,s}(\yvec)$ have the same sign
(or else $f_{i,s}(y)=0$, in which case $f_{i,s}$ is the constant function $0$),
so in both runs the machine entered the same state at
stage $s+1$, leaving the contents of all cells intact.
This completes the induction, and leaves us only to remark that
therefore, at stage $t$, the run of $M$ on input $\xvec$
must also have halted, with $\la f_{0,t}(\xvec),\ldots,f_{n,s}(\xvec)\ra$
in its cells as the output.
\end{Proof}

(If our BSS machines were allowed to compare the contents
of two cells under $=$ or $<$, as is standard, then
our set $F$ would have to consist of all nonconstant differences
$(f_{i,s}-f_{j,s})$.  The proof would still work, but the
method above is simpler.)

Lemma \ref{lemma:epnbhd} provides quick proofs of several
known results, including the undecidability of every
proper subfield $F\subset\R$.

\begin{cor}
\label{cor:decidable}
No BSS-decidable subset $S\subseteq \R^n$ can be both dense
and co-dense in $\R^n$.
\end{cor}
\begin{Proof}
Since the characteristic function $\chi_S$ is BSS-computable,
say by a machine with parameters $\zvec$,
Lemma \ref{lemma:epnbhd} shows that for every $\yvec\in\R^n$ with coordinates
algebraically independent over $\zvec$, $\chi_S$ is constant
in some neighborhood of $\yvec$.
\end{Proof}
Indeed, the same proof shows that any BSS-computable
total function with discrete image must be constant on each
of the $\ep$-balls given by Lemma \ref{lemma:epnbhd}.
\begin{cor}
\label{cor:bdry}
Define the boundary of a subset $S\subseteq\R^n$
to be the intersection of the closure of $S$ with the
closure of its complement.  If $S$ is BSS-decidable,
then there is a finite tuple $\zvec$ such that
every point on the boundary of $S$ has coordinates algebraically
dependent over $\zvec$.  In particular, if $M$ computes
$\chi_S$, then its parameters may serve as $\zvec$.
\end{cor}
\begin{Proof}
This is immediate from Lemma \ref{lemma:epnbhd}.
\end{Proof}

Of course, Corollaries \ref{cor:decidable} and \ref{cor:bdry}
have been deduced long since from other known results,
in particular from the Path Decomposition Theorem described in \cite{BCSS}.
We include them here because of the simplicity of these proofs,
and because they introduce the methods to be used in the following sections.

\section{Application to Algebraic Numbers of Degree $2$}
\label{sec:quadratic}

We know that $\A_{\leq 1}\not\leq_{BSS}\A_{\leq 0}$,
since $\A_{\leq 0}=\emptyset$ and $\A_{\leq 1}=\Q$
and it is already known (and seen again in Corollary
\ref{cor:decidable} above) that $\Q$ is BSS-undecidable.
To introduce our main result, we prove the corresponding
result one level further up.  The proof constitutes
a simple introduction to the method we used in the abstract
\cite{CKM10} to prove the full Theorem \ref{thm:general}.
In the next section, we will give a separate new proof
of Theorem \ref{thm:general}, more elegant
than the first one but less transparent,
especially for readers not expert in field theory,
who may prefer to look up the proof in \cite{CKM10}.

\begin{thm}
\label{thm:quadratic}
$\A_{\leq 2}\not\leq_{BSS}\A_{\leq 1}$.
\end{thm}
\begin{Proof}
Suppose that $M$ is an oracle BSS machine with real parameters $\zvec$,
such that $\M^{\Q}$ computes the characteristic function of $\A_{\leq 2}$.
(Of course $\A_{\leq 1}$ is just $\Q$ itself.)
Fix any $y\in\R$ which is transcendental over the field $Q=\Q(\zvec)$,
and run $M^{\Q}$ on input $y$.  Of course, this computation
must halt after finitely many steps and output $0$.  As in the proof
of Lemma \ref{lemma:epnbhd}, we set $F$ to be the finite set of
all nonconstant rational functions $f\in Q(Y)$ such that $f(y)$
appears in some cell during this computation.  Again, there is an
$\ep >0$ such that all $x$ within $\ep$ of $y$ satisfy
$f(x)\cdot f(y)>0$ for all $f\in F$.  However, it is no longer
sufficient for us to run $M^{\Q}$ on an arbitrary $x\in B_\ep(y)\cap\A_{\leq 2}$,
because such an $x$ might lie in $\Q$, or might have
$f(x)\in\Q$ for some $f\in F$, and in this case the computation
on input $x$ might ask its oracle whether $f(x)\in\Q$ and
would then branch differently from the computation on input $y$.
(Of course, for all $f\in F$, $f(y)\notin\Q$, since $f(y)$ must be
transcendental over $\Q$ for nonconstant $f$.)
So we must establish the existence of some $x\in B_\ep (y)\cap\A_{\leq 2}$
with $f(x)\notin\Q$ for all of the (finitely many) $f\in F$.
Of course, we do not need to give any effective procedure
which produces this $x$; its existence is sufficient.

For each $f\in F$, choose $g,h\in Q[Y]$ with $f=\frac{g}{h}$.
Thus $f(x)=a$ iff $g(x)-ah(x)=0$.
In game-theoretic terms, the opponent first chooses the functions in $F$,
after which we choose $x$, and then his functions pick out the value
of $a$, based on our $x$.  So we must ensure that our choice of $x$
makes $0\neq g(x)-ah(x)$ for every $a\in\Q$.



We need the following lemma from calculus.
\begin{lemma}
\label{lemma:deriv}
If $f\in\R(X)$ with $b\in\dom{f}$,
and there are positive values of $v$ arbitrarily close
to $0$ such that $f(b+v)=f(b-v)$, then $f'(b)=0$.
%
\qed\end{lemma}

Given the collection $F$, we fix some $b\in\Q$ such that
$|y-b|<\frac{\ep}{2}$ and such that $b$ lies in the
domain of every $f\in F$ and all $f'(b)\neq 0$.
Since each $f\in F$ is differentiable and
nonconstant, each $f$ rules out only finitely many
values, and $F$ is finite, so such a $b$ must exist.
Now Lemma \ref{lemma:deriv} makes it clear
that for some sufficiently small $\delta>0$,
every $u\in\Q$ with $0<\sqrtu<\delta$ satisfies
$f(b+\sqrtu)\neq f(b-\sqrtu)$ for every $f\in F$.
So fix $x=b+\sqrtu$ for some $u\in\Q$
with $0<\sqrtu<\min(\delta,\frac{\ep}{2})$, for which $\sqrtu\notin Q$.
(The finitely generated field $Q$ cannot contain
every $\sqrtu$ in this interval, as every subfield of a finitely generated field
is itself finitely generated.  For a proof, see
\cite[Thm.\ 3.1.4, p.\ 82]{N93}.)  Thus $|x-y|<\ep$
and all $f\in F$ satisfy $f(b+\sqrtu)\neq f(b-\sqrtu)$.
We write $\xbar=b-\sqrtu$ for the conjugate
of $x$ over $\Q$.

Now let $p(X) = X^2-2bX+(b^2-u)$, which is
the minimal polynomial of $x$ and of $\xbar$, over $Q$
as well as over $\Q$.  For each $f\in F$, we apply the division algorithm:
$$ f(X) =\frac{g(X)}{h(X)}=\frac{q_g(X)\cdot p(X)+r_g(X)}{q_h(X)\cdot p(X)+r_h(X)}$$
with $r_g(X)$ and $r_h(X)$ both linear polynomials.
We write $r_g(X)=g_1X+g_0$ and $r_h(X)=h_1X+h_0$,
with all coefficients in $Q$.
Now $f(x)=\frac{q_g(x)\cdot 0+r_g(x)}{q_h(x)\cdot 0+r_h(x)}=\frac{r_g(x)}{r_h(x)}$,
and likewise $f(\xbar)=\frac{r_g(\xbar)}{r_h(\xbar)}$.
Since $f(x)\neq f(\xbar)$, this shows that $\frac{r_g(X)}{r_h(X)}$
cannot be constant, so $r_g(X)$ is not a scalar multiple of $r_h(X)$.

Suppose that $a=f(x)=\frac{g(x)}{h(x)}$ lies in $Q$.  Then
$$ g_1\cdot (b+\sqrtu) +g_0=r_g(x)=g(x)=ah(x)=ar_h(x)=ah_1\cdot (b+\sqrtu)+ah_0,$$
and this equation can be re-expressed as
$$ (g_1b+g_0-ah_1b-ah_0)+\sqrtu(g_1-ah_1)=0.$$
Here both expressions in parentheses lie in $Q$,
but we chose $u$ with $\sqrtu\notin Q$, and so
$$ g_1b+g_0 = a(h_1b+h_0)\hspace{1cm}\text{and}\hspace{1cm}g_1=ah_1.$$
But this immediately shows that
$r_g(X) = g_1X+g_0=ah_1X+ah_0 = ar_h(X)$,
contradicting the statement above that $r_g(X)$
is not a scalar multiple of $r_h(X)$.

With this contradiction, we see that $f(x)\notin \Q$
(and indeed $f(x)\notin Q$).  Since this holds for all $f\in F$,
and since $|x-y|<\ep$, it is now clear 
(just as in Lemma \ref{lemma:epnbhd}) that the computation
$M^{\Q}(x)$ follows the exact same path as $M^{\Q}(y)$,
and outputs the same answer.  However, $y\notin\A_{\leq 2}$
since $y$ was transcendental over $Q$, whereas $x=b+\sqrtu\in\A_{\leq 2}$.
Thus the machine $M$ with oracle $\Q$ did not
compute the characteristic function of $\A_{\leq 2}$.
\end{Proof}

\section{Application to Algebraic Numbers in General}
\label{sec:general}

\begin{thm}
\label{thm:general}
For every $d>0$, $\A_{\leq d}\not\leq_{BSS}\A_{\leq d-1}$.
In particular, $\A_{=d}\not\leq_{BSS}\A_{\leq d-1}$.
\end{thm}
\begin{Proof}
The two statements in the theorem are equivalent,
because $\A_{\leq d}\equiv_{BSS}\A_{\leq d-1}\oplus\A_{=d}$
(where $A\oplus B=\set{\la 0,\avec\ra}{\avec\in A}
\cup\set{\la 1,\bvec\ra}{\bvec\in B}$).
We prove the latter.  

As usual, suppose that $M$ is an oracle BSS machine
which, given oracle $\A_{\leq d-1}$, computes $\A_{=d}$.
Let $\zvec$ be the finite tuple of real parameters
used by $M$, and set $Q=\Q(\zvec)$, the field
generated by these parameters.  Then its algebraic
portion $Q\cap\Qbar$ is also finitely generated,
since subfields of finitely generated fields are finitely generated,
(see \cite[Thm.\ 3.1.4, p.\ 82]{N93}).  Being an algebraic extension,
$Q\cap\Qbar$ thus has finite dimension over $\Q$, so
$Q$ cannot contain $d$-th roots of all prime numbers.  
Theorem 9.1 from Chapter 6 of \cite{L93} shows that
there exists a prime whose real $d$-th root $\alpha$
satisfies $[Q(\alpha):Q]=d$.
We also fix any real number $y$ transcendental over $Q$,
and let $F$ be the set of all nonconstant rational
functions $f\in Q(X)$ such that $f(y)$ appears
in some cell during the run of $M$ on input $y$
with oracle $\A_{\leq d-1}$.  By assumption this
run halts and outputs $0$, so $F$ is a finite set.
As before, we fix $\ep>0$ such that
$f(x)\cdot f(y)>0$ for all $x\in B_\ep (y)$.
Fix any $b\in\Q$ with $|b-y|<\frac{\ep}{2}$
and with $f'(b)\neq 0$ for all $f\in F$.
(Each such $f$ is a nonconstant rational function,
and so $f'(X)$, being rational and nonzero, can
only have finitely many roots.)

Now there exist $d$ distinct embeddings $\sigma_j:\Q(\alpha)\to\Qbar$,
with $j=1,\ldots,d$.  With $Q$ and $\Q(\alpha)$ linearly
disjoint, each $\sigma_j$ extends to an embedding
$\sigmabar_j:Q(\alpha)\to\overline{Q(\alpha)}$,
with $\sigmabar_j\res Q$ being the identity,
and these are all the embeddings (over $Q$)
of $Q(\alpha)$ into its algebraic closure.

\begin{lemma}
\label{lemma:roots}
In this situation, let $S$ be the set of
rational numbers $c$ such that $\beta_c=f(b+c\alpha)$
has degree $<d$ over $Q$.  Then $S$ is finite.
\end{lemma}
\begin{Proof}
For any $c\in S$, we have $Q(\beta_c)\subsetneq Q(\alpha)$,
with inclusion because $\beta_c=f(b+c\alpha)$ and $f\in Q(X)$,
and without equality because $[Q(\beta_c):Q]<d=[Q(\alpha):Q]$.
Hence some $\sigmabar_{j(c)}$ embeds $Q(\alpha)$ into its algebraic
closure and equals the identity on $Q(\beta_c)$ but not on $Q(\alpha)$.
But $\alpha$ has only $d$ conjugates over $Q$, so if $S$
were infinite, it would contain an infinite subset $S'$
for which one particular embedding $\sigmabar=\sigmabar_j$
would have $\sigmabar(\beta_c)=\beta_c$ for all $c\in S'$, but
$\sigmabar(\alpha)\neq\alpha$.

Let $\gamma=\frac{\sigmabar(\alpha)}{\alpha}\neq 1$.
Now for $c\in S'$,
$$ f(b+c\alpha)=\beta_c=\sigmabar(\beta_c)=\sigmabar(f(b+c\alpha))
=f(b+c\cdot\sigmabar(\alpha)).$$
So the equation $f(b+X)=f(b+\gamma X)$ holds whenever $X=c\alpha$
with $c\in S'$.  Since $S'$ is infinite, this must be an identity
of rational functions, and by differentiating it (and recalling
that $\gamma\neq 1$), we see that $f'(b)=0$,
contradicting our choice of $b$.
\end{Proof}

So we may fix some positive rational $c<\frac{\ep}{|2\alpha|}$
such that for all $f\in F$, $f(b+c\alpha)$ has degree $\geq d$ over $Q$.
(In fact, this degree must then equal $d$,
since $f(b+c\alpha)\in Q(\alpha)$.)
Let $x=b+c\alpha$.  Then $|y-x|\leq |y-b|+|b-x|<\frac{\ep}{2}+\frac{\ep}{2}=\ep$,
so that for all $f\in F$, $f(x)$ and $f(y)$ have
the same sign and both lie outside of the oracle set $\A_{\leq d-1}$.
Therefore the computation of $M$ on input $x$
with oracle $\A_{\leq d-1}$ halts and outputs $0$,
yet $x\in\A_{=d}$, so this machine does
not decide the set $\A_{=d}$.
\end{Proof}

Thus we have a positive answer to Problem \ref{mzq}.
Meer and Ziegler also posed a similar problem:  whether
$\A_{\leq d}\equiv_{BSS}\A_{=d}$ for every $d$.  They had noted that
this holds when $d\leq 2$, but in fact it holds for no other $d$ than those.
Since $\A_{\leq d}\equiv_{BSS}\A_{\leq d-1}\oplus\A_{=d}$, the problem
essentially asks whether $\A_{\leq d-1}\leq_{BSS}\A_{=d}$.
In the next section, Theorem \ref{thm:full} will use the same
technique as Theorem \ref{thm:general} to show that for $d-1>0$,
$\A_{=d-1}\leq_{BSS}\A_{=d}$ iff $(d-1)$ divides $d$.

\section{Results on Sets of Degrees}
\label{sec:hopeful}

Now we create more general versions, mostly along the same
lines as Theorem \ref{thm:general}.  To make the notation as
powerful as possible, we extend it:  for any subset $S\subseteq\omega$,
we will write $\A_S = \cup_{d\in S}\A_{=d}$, the set of all algebraic
real numbers whose degrees over $\Q$ lie in $S$.
\begin{lemma}
\label{lemma:semidecidable}
For every $S\subseteq\omega$, $\A_S$ is a BSS-semidecidable set,
and the semidecision procedure is uniform in one real parameter for $S$.
\end{lemma}
\begin{Proof}
The BSS machine $M$ with range $\A_S$ has one real parameter
$z_S=\sum_{n\in S} 2^{-n}$, whose binary representation forms
a code for the set $S\subseteq\omega$.
From this parameter, given any $n$, the machine can determine
whether or not $n\in S$.  On input $x$, the machine searches
through all irreducible $h(X)\in\Q[x]$ until it finds one with $h(x)=0$.
Then it uses $z_S$ to decide whether $\deg{h}\in S$,
and halts only on a positive answer.
(It was known as far back as 1882, in the work \cite{K1882}
of Kronecker, that there is a decision procedure for
irreducibility of polynomials in $\Q[X]$.)
\end{Proof}

Our first result on these BSS-semidecidable sets is immediate.
\begin{lemma}
\label{lemma:SinT}
If $S\subseteq T\subseteq\omega$, then $\A_S\leq_{BSS}\A_T$.
\end{lemma}
\begin{Proof}
With an $\A_T$ oracle, a BSS
machine can decide whether an input $x$ lies in $\A_T$.
If it does not (and $S\subseteq T$), then $x\notin\A_S$;
whereas if it does, then one merely searches for the minimal
polynomial of $x$ in $\Q$, and checks whether its degree
lies in $S$.  (For this purpose, the machine needs the
parameter $z_S$ from Lemma \ref{lemma:semidecidable}.)
\end{Proof}

\begin{thm}
\label{thm:full}
For every $d>0$ in $\omega$ and every set $S\subset \omega$ with
$S\cap d\Z=\emptyset$, $\A_{=d}\not\leq_{BSS}\A_{S}$.
\end{thm}
\begin{Proof}
Essentially the same construction as in Theorem \ref{thm:general} applies,
and we use the same notation.
In addition to the argument given there,
we must show that for every $f\in F$, $f(x)$ cannot lie in the new oracle set $\A_S$,
as opposed to $\A_{\leq d-1}$, which was the oracle used in Theorem
\ref{thm:general}.  For this purpose, a few revisions are needed.
First, with $Q$ still denoting the field generated by the parameters
of $M$, we let $K=Q\cap\overline{\Q}$, and let $F_K=F\cap K(X)$ be the set
of nonconstant functions in $K(X)$ appearing in the computation
by $M^{\A_S}$ on the transcendental input $y$.  Now
$K$ is a finite algebraic extension of $\Q$ within $\R$,
and we let $L$ be the Galois extension of $\Q$ generated
by $K$ within the complex field $\C$.  Moreover,
we will choose our $x$ to have degree $d$ not
only over $\Q$, and not only over $Q$, but also over $L$.
Since $L$ and $Q$ are both finitely generated, this can be done.

Now let $f\in F$ and consider $a=f(x)$.
First, if $f\notin F_K$, then every expression of $f(X)$
involves transcendentals over $\Q$.  Suppose
that $f(X)=f^*(t_1,\ldots,t_m,X)$,
where $f^*=\frac{g^*}{h^*}\in K(T_1\ldots,T_m,X)$ and $\{ t_1,\ldots,t_m\}$
(which we write as $\{\tvec\}$) is algebraically independent over $\Q$.
If $a=f(x)=f^*(\tvec,x)$ lies in $\A$, then
$g^*(\tvec,x)=ah^*(\tvec,x)$, so
all coefficients in $g^*(\Tvec,x)-ah^*(\Tvec,x)$
are zero, which can happen for only finitely many values of $x$
(since $f$ is nonconstant).  So, in the revised proof,
we make sure to avoid those
finitely many values of $x$ (for each $f\in F-F_K$) when
making our choice of $x$ close to $y$.

Next we suppose that $a=f(x)$ for an $f\in F_K$.
The same proof as in Theorem \ref{thm:general} shows that
$L(a)$ cannot be a proper subfield of $L(x)$.
(The argument there for $Q$ goes through now with
$Q$ replaced by $L$.)
Now, however, it is not guaranteed that $a$ has degree $<d$ over $L$,
and so it is possible that $L(a)=L(x)$.  In this case,
we know that $[L(a):L]=[L(x):L]=d$, and so $a$ has minimal
polynomial $q(X)\in L[X]$ of degree $d$.  If $a$ is transcendental
over $\Q$, then of course $a\notin\A_S$; so assume $a$ is algebraic,
and let $p(X)\in\Q[X]$ be the minimal polynomial of $a$ over $\Q$.
Then $p(X)$ is just the product of $q(X)$ with several images
of $q(X)$ under automorphisms of $L$.  Specifically, if $E$ is
the subfield of $L$ generated by the coefficients of $q(X)$, then
$$ p(X) =\Pi_{\sigma\in G}~q^\sigma (X),$$
where $G$ is a set of representatives for the right cosets
of $\Gal{L}{E}$ in $\Gal{L}{\Q}$, and where $q^\sigma(X)$
is the image of $q(X)$ under the map $\sigma$ on its coefficients.
(This formula for the minimal polynomial $p(X)$ of the roots of $q$
over the smaller subfield $\Q$
requires the extension $\Q\subseteq L$ to be Galois.
This is why we used the Galois extension $L$,
rather than just the fields $K$ or $Q$.)  
It follows that $\deg{p(X)}=d\cdot [E:\Q]$.
Since $S$ contains no nonzero multiples of $d$,
we see that $\deg{p(X)}\notin S$, and hence $a\notin\A_S$.
The rest of the proof then proceeds exactly as in Theorem \ref{thm:general}.
\end{Proof}

For reference in Theorem \ref{thm:SandT}, we note that
in the above proof, when $L(a)=L(x)$, we showed
that $[\Q(a):\Q] = d\cdot [E:\Q]$ is a multiple of $d$
by a factor $\leq [L:\Q]$.  Here all that was needed
was for $[\Q(a):\Q]$ to be a multiple of $d$, but there
we will need uniformity in the size of the multiple.

We will see below that the converse of
Lemma \ref{lemma:SinT} fails.  However, we can prove a substitute
for it, which yields the same principal results (Corollaries \ref{cor:lattice}
and \ref{cor:antichain}, below) that one would have derived from the converse.
\begin{thm}
\label{thm:lattice}
Let $P$ be the set of all prime numbers in $\omega$.
Then for all $S$ and $T$ in the power set $\P (P)$,
$A_S \leq_{BSS} A_T$ if and only if $S \subseteq T$.
\end{thm}
\begin{Proof}
The backward direction follows from Lemma \ref{lemma:SinT},
and the forward from Theorem \ref{thm:full},
since no element of $P$ divides any other element of $P$.
\end{Proof}
This allows us to show that the BSS-semidecidable degrees,
which form the analogue for BSS computation of the computably
enumerable Turing degrees, are a structure of significant complexity.
In \cite[Theorem 16]{MZ08}, Meer and Ziegler showed that there exist
uncountably many BSS-semidecidable degrees,
pairwise incomparable with each other, and Corollary \ref{cor:lattice}
(below) could also be proven using their sets $\Q_{\sqrt{p}}$,
and unions of those sets, in place of the sets $\A_{=p}$ and their unions.
In the case of additive BSS-machines, some related results
appear in \cite{G08}.

\begin{cor}
\label{cor:lattice}
There exists a subset $\mathcal{L}$ of the BSS-semidecidable
degrees such that as partial orders, $(\mathcal{L},\leq_{BSS})$
is isomorphic to $(\mathcal{P}(\omega),\subseteq)$.
\end{cor}
\begin{Proof}
We have $(\mathcal{P}(\omega),\subseteq)\cong(\mathcal{P}(P),\subseteq)$,
and Theorem \ref{thm:lattice} shows that the latter partial order
embeds into the BSS-semidecidable degrees via the
map $S\mapsto\A_S$.
\end{Proof}

In Corollary \ref{cor:lattice},
the isomorphism respects partial orders, but it is open
whether it maps meets and joins within the lattice
$(\mathcal{P}(\omega),\subseteq)$ to meets and joins
within the upper semilattice of the BSS degrees,
or within the sub-upper semilattice of the
BSS-semidecidable degrees.

By adding the following elementary set-theoretic fact to Corollary \ref{cor:lattice}
we create a different proof of \cite[Thm.\ 16]{MZ08}.

\begin{lemma}[Folklore]
\label{lemma:settheory}
There exists a collection
$\D\subseteq\P(\omega)$ of cardinality $2^\omega$ satisfying:
$$ (\forall A\in\D)(\forall B\in\D)[A\subseteq B\implies A=B].$$
\end{lemma}
\begin{Proof}
Recall that for $A,B\subseteq\omega$,
$A\oplus B=\set{2n}{n\in A}\cup\set{2n+1}{n\in B}$.
Let $\D=\set{S\oplus S^C}{S\in\P(\omega)}$,
where $S^C$ is the complement of $S$.
\end{Proof}

\begin{cor}[first shown in \cite{MZ08}]
\label{cor:antichain}
There exists an antichain (under $\leq_{BSS}$)
of cardinality $2^\omega$ within the BSS-semidecidable degrees.
\qed\end{cor}


The remainder of this section is devoted to the question
of when we can have $\A_S\leq_{BSS}\A_T$ without
having $S\subseteq T$.
The restriction on $S$ in Theorem \ref{thm:full} is that it does not
contain any multiples of $d$.  This leaves open the full
converse of Lemma \ref{lemma:SinT}, and it is natural to
conjecture that $\A_S\leq_{BSS}\A_T$
if and only if $S\subseteq T$.  The authors were surprised
to find that this conjecture fails, and moreover,
that it can fail even for finite sets $S$.
To introduce this failure, we give a specific example,
which builds on a technique introduced by Meer
and Ziegler in \cite[Lemma 17]{MZ08}, in which they
proved that $\A_{=1}\leq_{BSS}\A_{=2}$.

\begin{prop}
\label{prop:26}
$\A_{=2}\leq_{BSS}\A_{=6}$.
\end{prop}
\begin{Proof}
The machine for this reduction, with oracle $\A_{=6}$,
is simple.  It has $\sqrt[3]2$ as a parameter,
and on input $x$, it asks its oracle whether
$x+\sqrt[3]2\in\A_{=6}$.  If not, then it outputs
``No'' immediately; while if so, then it
searches through the irreducible polynomials
in $\Q[X]$ until it finds the minimal polynomial
$h(X)$ of $x$, and then outputs ``Yes'' if $h$
has degree $2$ and ``No'' otherwise.

To see that this procedure computes $\A_{=2}$ correctly,
we consider the algebraicity of the input $x$.
If $x\notin\A$, then $x+\sqrt[3]2$ is certainly
also transcendental, hence $\notin\A_{=6}$,
and the machine outputs ``No.''  If $x\in\A_{(\omega-\{2\})}$, then
the machine must output ``No'': either immediately,
if $x+\sqrt[3]2\notin\A_{=6}$, or else after finding
the minimal polynomial of $x$.
Finally, if $x\in\A_{=2}$,
then we claim that $x+\sqrt[3]2\in\A_{=6}$,
and so the machine goes into its search for the minimal
polynomial of $x$, and finds that $x\in\A_{=2}$.

To see that $x+\sqrt[3]2\in\A_{=6}$ when $x\in\A_{=2}$,
let $F$ be the field $\Q(x+\sqrt[3]2)$.  Then
$F\subseteq\Q(x,\sqrt[3]2)$, and we show that indeed equality holds.
First, $F(x)=\Q(x,\sqrt[3]2)$, and so $\Q(x,\sqrt[3]2)$
must have degree either $1$ or $2$ over $F$.
But also $F(\sqrt[3]2)=\Q(x,\sqrt[3]2)$, so $\Q(x,\sqrt[3]2)$
must have degree either $1$ or $3$ over $F$.
The only consistent solution is that $F=\Q(x,\sqrt[3]2)$,
so $[F:\Q]=6$.  Thus $x+\sqrt[3]2\in\A_{=6}$ as required.
\end{Proof}

The only special aspect of the numbers $2$ and $6$ here
was that $\frac{6}{2}$ is an integer relatively prime to $2$.
\begin{prop}
\label{prop:pq}
Let $p$ and $q$ be any positive, relatively prime integers.
Then $\A_{=p}\leq_{BSS}\A_{=pq}$.
\end{prop}

\begin{Proof}
The exact same proof as in Proposition \ref{prop:26},
with $2$ replaced by $p$, $6$ by $pq$, and $\sqrt[3]2$
by any element of $\A_{=q}$ one may choose.
\end{Proof}

The preceding argument can be uniformized, via
a specific effective version of the Theorem of the Primitive Element.
The basic result was proven by Kronecker; we suggest
\cite[Lemma 17.12]{FJ86} for a description.
Here we simply adapt that proof to the BSS setting.
\begin{thm}[Effective Theorem of the Primitive Element,
after Kronecker]
\label{thm:PrimElt}
There is a BSS-computable function which,
given any two finite (possibly empty) tuples $\la x_1,\ldots,x_n\ra$
and $\la y_1,\ldots,y_m\ra$ of real numbers
such that $\{x_1,\ldots,x_n\}$ is algebraically independent
over $\Q$ and all $y_i$ are algebraic over $\Q(x_1,\ldots,x_n)$,
outputs a single real number $z$ such that
$\Q(\xvec,\yvec)=\Q(\xvec,z)$.
(This $z$ is called a \emph{primitive element}
for the field $\Q(\xvec,\yvec)$ over $\Q(\xvec)$.)
\end{thm}
\begin{Proof}
We use the procedure from \cite[Lemma 17.12]{FJ86},
with $K=\Q(\xvec)$, noting that this $K$ has a
(Turing-computable) splitting algorithm, uniformly in $n$,
so that our machine can begin by finding the minimal polynomial
of each $y_{i}$ over $K(y_1,\ldots,y_{i-1})$,
for $i=1,\ldots,m$.  It can also find the minimal polynomial
$f(Y)$ of the element $y=T_1y_1+\cdots +T_my_m$ over the function
field $K(T_1,\ldots,T_m)$, clearing denominators
so that $f\in K[\Tvec,Y]$.  As argued in \cite{FJ86},
for any $a_1,\ldots,a_m\in K$ such that
$\frac{\partial f}{\partial Y}(a_1,\ldots,a_m,a_1y_1+\cdots +a_my_m)\neq 0$,
the element $z=a_1y_1+\cdots +a_my_m$
generates $K(\yvec)=\Q(\xvec,\yvec)$ over $K$,
and our machine will output such a $z$.

For BSS machines, finding an appropriate tuple
$\la a_1,\ldots,a_m\ra$ requires the machine to enumerate
the elements of $K$, searching for such a tuple.
Recall that a set $S\subseteq\R^\infty$ is
\emph{BSS-countable} if there exists a BSS machine $M$
such that $S$ is the image of the set $\N$ under $M$.
Uniform BSS-countability in an input is then
defined in the natural way:  a collection $S_{\xvec}$
of sets, indexed by elements $\xvec$ of $\R^\infty$,
is uniformly BSS-countable in these tuples if there is
a BSS machine $M'$ such that for all indices $\xvec$,
$S_{\xvec}=\set{\text{outputs~}M'(\xvec,j)}{j\in\N}$.
It is clear that the fields $K=\Q(\xvec)$ above are
BSS-countable uniformly in $\xvec$, and so our machine
can search through elements of $K^m$ to find the requisite tuple.
\end{Proof}

\begin{prop}
\label{prop:relprime}
Let $S$ and $T$ be subsets of the positive integers.
Suppose that for some absolute constant $N$ and each $d \in S$,
there is a positive integer $n_d \le N$ and prime to $d$ such that
$dn_d \in T$.  Then $\A_S \le_{\rm BSS} \A_T$.
\end{prop}

\begin{Proof}
For $n=1,\ldots,N$, fix parameters $z_n=\sqrt[n]2\in\R$, so each
$\Q(z_n)$ is an extension of degree $n$ over $\Q$.
On input $x$, our machine computes a primitive element
$a_n$ for each of the fields $\Q(x,z_n)$, using
Theorem \ref{thm:PrimElt}, and checks whether any of the $N$
elements $a_n$ lies in the oracle set $\A_T$.
If any one does, we know $x$ is algebraic, so we search for the minimal
polynomial of $x$ over $\Q$ and output ``yes'' or ``no'' depending
on whether or not its degree is in $S$.  

Suppose that no $a_n$ lies in $\A_T$.
In this case the machine program outputs ``no.''
Since this clearly is the correct answer when
$x$ is transcendental, we consider $x$ algebraic
of some degree $d$ over $\Q$.
If $d \in S$, let $m = n_d$.  By assumption, $\gcd(m,d) = 1$,
so $\Q(x)\cap\Q(z_m)=\Q$, forcing
$[\Q(a_m):\Q]=[\Q(x,z_m):\Q] = dm\in T$,
and therefore $a_m\in\A_T$.
\end{Proof}

Next we show how to remove the assumption of relative primality
for a finite set $S$.

\begin{prop}
\label{prop:allpq}
Let $p$ and $r$ be any positive integers.
Then $\A_{=p}\leq_{BSS}\A_{=r}$ if and only if $p$ divides $r$.
\end{prop}
\begin{Proof}
The ``only if'' direction follows from Theorem \ref{thm:full}.
For the ``if'' direction, let $q=\frac{r}{p}$.
We start with parameters $z_0,\ldots, z_n$,
where $n=2^p-2$ and $z_i$ is the
positive real $q$-th root of the $i$-th prime number:
$z_0=\sqrt[q]2, z_1=\sqrt[q]3,\ldots$.
Notice that then $\Q(z_i)\cap\Q(z_j)=\Q$ for all
$i\neq j$.

On input $x$, this machine asks its $\A_{=r}$-oracle
whether any of the elements $a_{ij}=z_i+jx$
lies in $\A_{=r}$, where $i$ and $j$ are
integers satisfying $0\leq i\leq n$
and $1\leq j\leq q(p-1)+1$.
If not, it outputs ``No,'' while if $(z_i+x)\in\A_{=r}$,
it finds the minimal polynomial of $x$ over $\Q$ and
outputs ``Yes'' or ``No'' depending on whether that
polynomial has degree $p$.

Once again, this machine clearly outputs ``No''
whenever $x$ (and hence all $a_{ij}$)
are transcendental, and clearly outputs
the correct answer whenever it actually searches
for the minimal polynomial of $x$ (because it
only does so when $x$ is algebraic).
The crucial situation is that in which $x$ is algebraic
and no $a_{ij}$ lies in $\A_{=r}$, so that
the output ``No'' comes immediately.
We must show that no such $x$ can lie in $\A_{=p}$.

So suppose $x\in\A_{=p}$.  Then the minimal polynomial
$h(X)\in\Q[X]$ of $x$ has exactly $2^p-2$ nontrivial
factors in $\overline{\Q}[X]$, corresponding to the
proper nonempty subsets of the set of all $p$ roots of
$h(X)$ in the algebraic closure $\overline{\Q}$.
Now $h$ has no proper factors in $\Q[X]$, hence any proper factor
of $h$ in one $\Q(z_i)[X]$ cannot lie in any other $\Q(z_j)[X]$,
(since $\Q(z_i)\cap\Q(z_j)=\Q$), leaving at least one
$i\leq n$ such that $\Q(z_i)[X]$ contains no proper factors 
of $h$ at all.  We fix this $i$, and note that then
the field $\Q(x,z_i)$ has degree $p$ over $\Q(z_i)$
(since $h(X)$ remains irreducible in $\Q(z_i)[X]$),
hence has degree $pq=r$ over $\Q$.

Now we claim that for this $i$, there exists at least
one $j\in\omega$ with $1\leq j\leq q(p-1)+1$ such that
$a_{ij}=z_i+jx$ is a primitive generator of $\Q(x,z_i)$.
This $a_{ij}$ must then lie in $\A_{=r}$,
so this will complete the proof.
The necessary result follows from the proof
given in \cite[\S 6.10]{vdW70} of the Theorem
of the Primitive Element.  Specifically,
it is shown there (with $\Delta=\Q$ being separable)
that $\Q(x,z_i)$ has a primitive generator
of the form $z_i+cx$, with $c\in\Q$, and indeed
that there are at most $q(p-1)$ values of $c$ in $\Q$
for which $z_i+cx$ \emph{fails} to be a primitive generator.
Since we tested $q(p-1)+1$ different values of $j$,
at least one $a_{ij}$ does generate $\Q(x,z_i)$,
hence has degree $r$ over $\Q$ as required.
\end{Proof}

We can generalize Proposition \ref{prop:allpq}
to finitely many degrees.
\begin{prop}
\label{prop:finiteS-T}
For any subsets $S$ and $T$ of $\omega$,
if $(S-T)$ is finite and for every $p\in S-T$,
there exists an integer $q>0$ such that
$pq\in T$, then $\A_S\leq_{BSS}\A_T$.
\end{prop}
\begin{Proof}
Let $S-T=\{ p_1,\ldots,p_k\}$, so that
$$\A_S\equiv_{BSS}\A_{S\cap T}\oplus\A_{=p_1}\oplus\cdots\oplus\A_{=p_k}
=\set{\la i,x\ra}{(i=0~\&~x\in\A_{S\cap T})\text{~or~}(x\in\A_{=p_i})}.$$
(The set on the right is clearly the least upper bound under $\leq_{BSS}$
of $\A_{S\cap T}$ and the sets $\A_{=p_i}$.)
Now $\A_{S\cap T}\leq_{BSS}\A_T$ by Lemma
\ref{lemma:SinT}, and by assumption,
for each $p_i$, there is some $q_i>0$ with
$p_iq_i\in T$, so that $\A_{=p_i}\leq_{BSS}\A_{=p_iq_i}\leq_{BSS}\A_T$,
using Proposition \ref{prop:allpq} and Lemma \ref{lemma:SinT}.
Hence $\A_S\leq_{BSS}\A_T$.
\end{Proof}

In the preceding construction, each element
of $(S-T)$ requires its own parameters in the
given BSS reduction.  However, Proposition
\ref{prop:relprime} showed that in certain cases
the reduction can be done uniformly,
and so $(S-T)$ need not be finite.
Next we show here that the uniformity in Proposition
\ref{prop:relprime} was essential.
\begin{thm}
\label{thm:SandT}
For sets $S,T\subseteq\omega$,
if $\A_S\leq_{BSS}\A_T$, then there exists
$N\in\omega$ such that all $p\in S$ satisfy
$\{ p,2p,3p,\ldots,Np\}\cap T\neq\emptyset$.
\end{thm}
\begin{Proof}
We prove the contrapositive,
assuming that there is no such $N$.  Suppose
an oracle BSS machine $M$ with parameters $\zvec$
computes $\chi_{\A_S}$ from oracle $\A_T$.
Let $Q=\Q(\zvec)$ as usual, and let $L$ be the
normal closure of $Q\cap\A$ within $\C$.
Thus $L$ is a Galois extension of $\Q$.
By assumption we may fix some $d\in S$ such that
$\{ d,2d,\ldots,[L:\Q]\cdot d\}\cap T=\emptyset$.
Running our usual argument with any $y\in\R$ transcendental
over $Q$, we get an $\ep$ and a finite set $F\subset Q(Y)$.
Now suppose $x\in\A_{=d}$ lies in $B_\ep(y)$ and has degree $d$
over $L$ as well as over $\Q$.  Then for every $f\in F$,
either $f(x)\notin\A$ (so $f(x)\notin\A_T$), or
$L(f(x))=L(x)$, or $L(f(x))$ is a proper subfield of $L(x)$.
But now, using the same argument as in Theorem
\ref{thm:full} for the choice of $x$, and referring to the
note at the end of the proof of that theorem,
we see that the case of a proper subfield may be avoided,
and that when $L(f(x))=L(x)$, the degree of $f(x)$ over $\Q$
equals $d\cdot [E:\Q]$.  Recall that $E$ was the subfield of $L$
generated by the coefficients of the minimal polynomial
of $f(x)$ over $L$.  Therefore $[E:\Q]\leq [L:\Q]$, and so
$d\cdot [E:\Q]$, the degree of $f(x)$ over $\Q$,
does not lie in $T$, by our choice of $d$.
Hence $f(x)\notin\A_T$, and the usual argument
then shows that the computation $M^{\A_T}(x)$
proceeds along the same path as for $y$, hence outputs
$0$, even though $x\in\A_{=d}\subseteq\A_S$.
\end{Proof}

For the converse of Theorem \ref{thm:SandT}, one would
need to uniformize the construction in Proposition \ref{prop:allpq},
along the lines of Proposition \ref{prop:relprime}.
We leave this question for another time.

\section{Countable Oracle Sets}
\label{sec:oracle}

It is natural to think of countability of a subset
$S\subseteq\R^\infty$ as a limit on the amount
of information which can be encoded into $S$.
This intuition requires significant restating before it can
be made into a coherent (let alone true) statement,
but we will give a reasonable version in this section.
So far, all our oracle sets have been of the form $\A_S$,
for various $S\subseteq\omega$, and so they have all been countable.
In \cite{MZ08}, it was asked whether there could exist a countable
set $C\subseteq\R^\infty$ such that the halting problem $\bH$
for BSS computation on $\R$ satisfies $\bH\leq_{BSS} C$.
We will show that the answer to this question is negative.
For a definition of $\bH$ in this context,
we refer the reader to \cite{MZ08}.
Since it is equiconsistent with \textbf{ZFC}
for the Continuum Hypothesis to be false,
we will make our arguments applicable to all infinite cardinals
$\kappa <2^\omega$, countable or otherwise.  Many of the results of
the present and subsequent sections were first announced in \cite{CKMCCA10}.

First, of course, every subset of $\R^\infty$ is BSS-equivalent
to its complement, and so countability and co-countability
impose the same restriction on information content.
Of course, many sets of size continuum, with equally large complements,
are quite simple:  the set of positive real numbers,
for example, is BSS-decidable, hence less complex than
the countable set $\A$.  So it is not possible to prove
absolute results relating cardinality and co-cardinality
(within $\R^\infty$) to BSS reducibility, but nevertheless,
we can produce theorems expressing the intuition
that countable sets are not highly complex in the BSS model.
This process will culminate in Theorem
\ref{thm:cardinality} below, but first we show that with a countable oracle,
one cannot decide the BSS halting problem $\bH$.
We conjecture that $\bH$ is not an upper bound on the degree of a countable set,
i.e.\ that such a set can still be BSS-incomparable with $\bH$,
but it certainly constitutes progress just to know that the upper
cone of sets above $\bH$ contains no countable sets.

\begin{thm}
\label{thm:haltingproblem}
If $C\subseteq\R^\infty$ is a set such that $\bH\leq_{BSS} C$,
then $|C|=2^\omega$.
\end{thm}
We note that by BSS-equivalence, these conditions also ensure
$|\R^\infty -C|=2^\omega$,
and ensure $|\R^m-C|=2^\omega$ whenever $C\subseteq\R^m$.

\begin{Proof}
Let $C\subseteq\R^\infty$ have cardinality $<2^\omega$,
and suppose that $M$ is an oracle BSS machine such that
$M^C$ computes the characteristic function of $\bH$.  We fix a program
code number $p$ for the program which takes inputs
$\la x_1,x_2\ra\in\R^2$, searches through nonzero polynomials $q$
in $\Q[Y_1,Y_2]$, and halts iff it finds one with $q(x_1,x_2)=0$.
Since the program coded by $p$ uses no real parameters,
$p$ may be regarded as a natural number, but in our
argument it can equally well be a tuple $\pvec$
from $\R^\infty$, with one or several real numbers
coding program parameters.  Then the elements of $C$, the
finitely many parameters $\zvec$ of $M$, and the parameters,
if any, in the program coded by $\pvec$ together generate
a field $E\subseteq\R$ which also has cardinality
$<2^\omega$, and so $\R$ is an extension of infinite
transcendence degree (indeed of degree $2^\omega$) over this $E$.
(Since $C\subseteq\R^\infty$, we need to be precise:
$E$ is generated by the coordinates $p_1,\ldots,p_j$ and $z_1,\ldots,z_k$
of the tuples $\pvec$ and $\zvec$, and the coordinates of each tuple in $C$.)
 
Now fix a pair $\la y_1,y_2\ra$ of real numbers algebraically
independent over $E$.  Hence $\la \pvec,y_1,y_2\ra\notin\bH$,
so $M^C$ on this input halts after finitely many steps
and outputs $0$.  We fix the finitely many
functions $f_{i,s}(\Yvec)\in E(Y_1,Y_2)$ such that $f_{i,s}(y_1,y_2)$
appears in the $i$-th cell at stage $s$ during this computation.
(The program code $\pvec\in E^\infty$ will stay fixed throughout
this proof, so we may treat it as part of the function $f_{i,s}$,
rather than as a variable.)
As usual, $F$ will be the set of those functions $f_{i,s}$ which
are not constants in $E$, and we fix an $\ep>0$
such that whenever $\la x_1,x_2\ra\in\R^2$ with $x_1\in B_\ep(y_1)$
and $x_2\in B_\ep(y_2)$, every $f\in F$ satisfies $f(x_1,x_2)\cdot f(y_1,y_2)>0$.
Write each $f\in F$ as a quotient $f=\frac{g}{h}$ with
$g,h\in E[Y_1,Y_2]$, and let $n$ be the greatest
degree of $Y_2$ in all of these finitely many polynomials $g$ and $h$.

Now choose $x_1\in\R$ to be transcendental over $E$
and within $\ep$ of $y_1$, and pick $x_2$
within $\ep$ of $y_2$ such that $x_2$ is 
algebraic over $\Q(x_1)$ but has degree $>n$ over $E(x_1)$.
For instance, let $x_1=y_1$ and $x_2=\sqrt[m]{x_1}+b$,
where $m>n$ is prime and $b\in\Q$ is selected to place
$x_2\in B_{\ep}(y_2)$.  The subfield $E(x_1)$ of $\R$
contains no nontrivial $m$-th roots of unity, nor any $m$-th roots
of $x_1$ (by our choices), so the Galois group of the splitting field of $(Y^m-x_1)$
over $E(x_1)$ is just the Galois group of its splitting field over $\Q(x_1)$,
which acts transitively on the roots.  Thus this polynomial is
irreducible over $E(x_1)$, and so $x_2$
has degree $m$ over $E(x_1)$.

Thus, for any $f\in F$, if $a=f(x_1,x_2)\in E$,
then $0=g(x_1,x_2)-ah(x_1,x_2)$.  Since $f$ is nonconstant,
$g$ is not a scalar multiple of $h$, and so $(g-ah)$ would then
be a nonzero polynomial in $E[Y_1,Y_2]$ of degree $\leq n$,
contradicting our choice of $x_2$.  Hence $f(x_1,x_2)\notin E$
for every $f\in F$.  But then the oracle computation
$M^C(\pvec,x_1,x_2)$ must follow the same path as $M^C(\pvec,y_1,y_2)$
and give the same output, namely $0$.  Since $\la \pvec,x_1,x_2\ra\in\bH$,
this proves that $M^C$ does not compute the characteristic function of $\bH$.
\end{Proof}

Indeed the preceding proof shows slightly more than was stated.
\begin{cor}
\label{cor:subfield}
If $C\subseteq\R^\infty$ is a set such that $\bH\leq_{BSS} C$,
then $\R$ has finite transcendence degree over the field
$K$ generated by (the coordinates of the tuples in) $C$,
and also has finite transcendence degree over the field
generated by the complement of $C$.
\end{cor}
\begin{Proof}
Given an oracle BSS machine $M$ which computes $\bH$ from oracle $C$,
let $E$ be the extension field $K(\zvec,\pvec)$, with $K$ as defined in the
corollary.  If $\R$ had transcendence degree $\geq 2$ over this $E$,
then the proof of Theorem \ref{thm:haltingproblem}
would go through:  we could choose $y_1,y_2\in\R$
algebraically independent over $E$, say with $y_1>0$,
and then let $x_1=y_1$ and $x_2=b+\sqrt[m]{x_1}$,
with $m>n$ prime, as in the proof, and
with $b\in\Q$ selected to put $x_2\in B_{\ep}(y_2)$.
But this would show that $M^C$ does not compute $\bH$.
So $\R$ has transcendence degree $\leq 1$ over
this $E$, and therefore is algebraic over $E(t)=K(t,\zvec,\pvec)$
for some $t\in\R$.

Since $C$ is BSS-equivalent to its complement,
the same proof applies to $(\R^\infty-C)$,
and also to $(\R^m-C)$ if $C\subseteq\R^m$.
\end{Proof}

As we consider the general case of a BSS computation
of the characteristic function $\chi_S$ of a set $S\subseteq\R$
using an oracle $C$ of infinite cardinality $\kappa<2^\omega$,
the following definition will be useful.
\begin{defn}
\label{defn:bicardinality}
A set $S\subseteq\R$ is \emph{locally of
bicardinality $\leq\kappa$} if there exist
two open subsets $U$ and $V$ of $\R$
with 
$|\R-(U\cup V)|\leq\kappa$
and $|U\cap S|\leq \kappa$ and
$|V\cap S^C|\leq \kappa$.
\end{defn}

If $\kappa<2^\omega$, then such $U$ and $V$ must be disjoint, since
$(U\cap V)$ is open with $|U\cap V|
\leq |U\cap S|+|V\cap\Sbar|\leq\kappa$.
So the definition roughly says that up to sets of size $\kappa$,
each of $S$ and $\Sbar$ is equal to an open subset of $\R$.
It is not equivalent to weaken
the requirement $|\R-(U\cup V)|\leq\kappa$ 
in Definition \ref{defn:bicardinality},
for instance by requiring that $|U\cap V|<2^\omega$.
For a counterexample, let $V=\emptyset$ and let $U$ be the complement
$S^C$ of the Cantor middle-thirds set $S$, which contains
all real numbers $x$ whose non-integer part
$x-\lfloor x\rfloor$ has a ternary expansion in
only $0$'s and $2$'s.  Thus
$U\cap S=V\cap\S^C=U\cap V=\emptyset$, yet this $S$ is not locally
of bicardinality $\leq\omega$ (nor $\leq$
any other $\kappa<2^\omega$),
as shown in full in Lemma \ref{lemma:Cantorbicard} below.

The \emph{local bicardinality of $S$} is the least cardinal
$\kappa$ such that $S$ is locally of bicardinality $\leq\kappa$.

The property of having local bicardinality $\leq\kappa$ does not appear
to us to be equivalent to any more easily stated property,
and we are not aware of it having been used (or even
stated) elsewhere in the literature.
The same definition in higher dimensions
completely loses its power: any connected component $U_0$
of $U$ must have boundary $\partial U_0$
with $U_0\cap\partial U=V\cap\partial U_0=\emptyset$,
since $U$ and $V$ are open and disjoint.  But then
$|\partial U_0|\leq |\R^n-(U\cup V)|\leq\kappa$,
which is feasible in $\R^1$ but not in higher dimensions,
unless $U$ or $V$ were empty or $\kappa=2^{\omega}$.
Thus, in $\R^n$ with $n>1$, every set of local bicardinality
$<2^{\omega}$ has either cardinality $<2^{\omega}$
or co-cardinality $<2^{\omega}$.
Nevertheless, within $\R^1$, this is exactly the condition
needed in our general theorem on cardinalities.

\begin{thm}
\label{thm:cardinality}
If $C\subseteq\R^\infty$ is an oracle set of
infinite cardinality $\kappa <2^\omega$, and
$S\subseteq\R$ is a set with $S\leq_{BSS} C$, 
then $S$ must be locally of bicardinality $\leq\kappa$.
The same holds for oracles $C$ of infinite co-cardinality
$\kappa<2^\omega$.
\end{thm}
\begin{Proof}
Again let $\zvec$ be the parameters used by the oracle
BSS machine $M$ which, given oracle $C$, computes $\chi_S$.
Then for any input $y\in\R$ transcendental over the subfield
$E$ of cardinality $\kappa$ generated by $\zvec$ and the individual
coordinates of all elements of $C$, there will again exist a finite set
$F_y\subseteq E(X)$ as above, and an $\ep>0$ such that $f(x)\cdot f(y)>0$
for all $x\in B_\ep (y)$ and $f\in F_y$.  For each such $y$,
let $B(y)$ be an open interval of length less than
the corresponding $\ep$, such that $B(y)$ contains $y$
and has rational end points.
Now if $x\in B(y)$ is also transcendental over $E$, then the
computation of $\chi_S(x)$ using this machine and the
$C$-oracle proceeds along the same path as the
computation for $y$, since $f(x)\notin E$ for all $f\in F_y$.
(Indeed, this would hold whenever $x\in B(y)$
has degree $>n$ over $E$, where $n$ is the maximum
degree of all numerators and denominators of elements of $F_y$.)
This shows that $\chi_S(x)=\chi_S(y)$
for all such $x$.  Since only $\kappa$-many elements
of $B(y)$ can be algebraic over the
size-$\kappa$ field $E$, it follows that either
$S\cap B(y)$ or $S^C\cap B(y)$
has size $\leq\kappa$.

Now if $t\in B(y_0)\cap B(y_1)$ is transcendental over $E$,
then $t$ follows the same computation path as both $y_0$ and $y_1$,
implying that $\chi_S(y_0)=\chi_S(y_1)$ whenever $B(y_0)\cap B(y_1)\neq\emptyset$,
and therefore that either $B(y_0)\cap S$ and $B(y_1)\cap S$
both have size $\leq\kappa$, or else $B(y_0)\cap\S^C$
and $B(y_1)\cap\S^C$ both have size $\leq\kappa$.
So when we set
$$ U=\bigcup\set{B(y)}{|S\cap B(y)|\leq\kappa}
\text{~~~and~~~} V=\bigcup\set{B(y)}{B(y)\not\subseteq U},$$
we will have $U\cap V=\emptyset$.  Here the
unions are over those $y\in\R$ transcendental over $E$
(as $B(y)$ is not defined for any other $y$),
and so the complement $\R-(U\cup V)$ is a subset
of the algebraic closure of $E$, which has size $\kappa$.
Moreover, being a union of open intervals $B(y)$ with rational
end points, $U$ in fact equals the union of countably
many such intervals, say $U=\cup_{i\in\omega} B(y_i)$
for some sequence $y_0,y_1,\ldots$.  Since each
$B(y_i)$ has intersection of size $\leq\kappa$ with $S$
(and since $\kappa\geq\omega$),
so does the entire union $U$.  Likewise
$|\S^C\cap V|\leq\kappa$, proving the theorem.


The claim about oracles of co-cardinality $\kappa$
follows from applying the same argument to the
oracle $(\R^\infty-C)$, which is BSS-equivalent to $C$.
If $C\subseteq\R^m$ for some $m$, then the same
holds of $(\R^m-C)$.
\end{Proof}

Notice that the set $S$ of smaller complexity must be a subset of $\R$,
whereas $C$ is allowed to contain tuples from $\R^\infty$.
We conjecture that to extend the theorem to sets
$S\subseteq \R^\infty$, we would need to allow
$\R^\infty-(U\cup V)$ to be a size-$\kappa$ union
of proper algebraic varieties defined over the field
generated by $C$.  
It is an open question (of interest only under
$\neg\textbf{CH}$) whether it is equivalent, for the purposes
of this conjecture and Theorem \ref{thm:cardinality},
to replace $|\R^\infty-(U\cup V)|\leq\kappa$
by $|\R^\infty-(U\cup V)|\leq\omega$ here
or in Definition \ref{defn:bicardinality}.

To understand that this theorem cannot readily be stated
using a simpler property than Definition \ref{defn:bicardinality},
consider the BSS-computable set
$$ S=\set{x\in (0,1)}{(\exists m\in\omega)~2^{-(2m+1)}\leq x-\lfloor x\rfloor\leq 2^{-(2m)}},$$
containing those $x\in (0,1)$ which have a binary expansion
with an even number of zeroes following the decimal point,
along with all translations of this set by integers.
The closure of $S$ is just $S\cup\Z$, but any open
set containing any integer $z$ would intersect
each of $S$ and $\S^C$ in $2^\omega$-many points.
So the theorem cannot require the complement
$\R^\infty-(U\cup V)$ to be finite, let alone empty.
With such tricks one can
create examples defying most conceivable simplifications
of Theorem \ref{thm:cardinality}.

In the next section we discuss the Cantor set, which
is often another useful counterexample in this vein.

\section{The Cantor Set}
\label{sec:Cantor}

As an example of a set of local bicardinality $2^{\omega}$,
we consider the Cantor set $\fC$, well known as
a set of measure $0$ within $\R$ which nevertheless
has cardinality $2^{\omega}$.  By definition, $\fC$ contains
all real numbers $x\in [0,1]$ having ternary expansions in
only $0$'s and $2$'s.  One usually views $\fC$
as the set of numbers in the unit interval $[0,1]$
which remain after $\omega$-many
iterations of deleting the open ``middle third'' of each interval
(starting with the middle third $(\frac13,\frac23)$ of $[0,1]$).
It is clear from this description that $\fC$ is co-semidecidable
in the BSS model:  even a Turing machine could enumerate
the end points of all those middle-third intervals to be deleted.
Hence $fC$ is $1$-reducible
to the complement $\bH^C$, forcing $\fC\leq_{BSS}\bH$.
The natural next question, whether $\bH\leq_{BSS} \fC$,
was settled by Yonezawa in \cite{Y08}, and we
thank the anonymous referee of \cite{CKMCCA10}
who pointed out the necessary result there.
\begin{thm}[Corollary 2.5 in \cite{Y08}]
\label{thm:Yonezawa}
The sets $\Q$ and $\fC$ are BSS-incomparable.
\qed\end{thm}
Since the BSS-semidecidable set $\Q$ must be $\leq_{BSS}\bH$,
this immediately answers the question:   $\bH\leq_{BSS} \fC$
would imply $\Q\leq_{BSS}\fC$, contradicting Theorem \ref{thm:Yonezawa}.
\begin{cor}
\label{cor:Yonezawa}
$\bH\not\leq_{BSS}\fC$.
\qed\end{cor}

Now we consider the local bicardinality of $\fC$.
The next lemma, combined with Theorem \ref{thm:cardinality},
immediately proves that $\fC$ is not BSS-decidable, nor even
BSS-semidecidable, in any oracle of size $<2^\omega$.
\begin{lemma}
\label{lemma:Cantorbicard}
The Cantor set $\fC$ has local bicardinality $2^{\omega}$.
\end{lemma}
\begin{Proof}
Suppose $\fC$ were locally of bicardinality $\leq\kappa <2^{\omega}$.
Then we would have open disjoint sets $U$ and $V$
satisfying Definition \ref{defn:bicardinality}, and $\fC$, having size
$2^{\omega}$, would have to intersect $V$ in some point $x$, since
$$ \fC-V\subseteq (U\cap \fC)\cup (U\cup V)^C$$
and the right-hand side has size $\leq\kappa$.  The open set $V$
would then contain an $\ep$-ball around $x$.
However, every open interval around $x$
intersects each of $\fC$ and $\fC^C$ in $2^{\omega}$-many points.
(To see this, just consider all $y$ whose ternary expansions match
that of $x$ for sufficiently many places to lie within that interval.)
Therefore $|V\cap\fC^C |=2^{\omega}$, yielding a contradiction.
\end{Proof}
\begin{cor}
The Cantor set $\fC$ is not BSS-semidecidable below $\A$,
or below any other oracle of cardinality $<2^{\omega}$.
\end{cor}
\begin{Proof}
This simply means that no function which is BSS-computable in
the oracle $\A$ can have $\fC$ as its domain.  Indeed,
if it did, then $\fC\leq_{BSS}\A$, since $\fC$ and $\fC^C$ would both be
$\A$-semidecidable.  By Lemma \ref{lemma:Cantorbicard}, we know that
$\fC$ has local bicardinality $2^\omega$, so that by Theorem
\ref{thm:cardinality} we have $|\A| \geq 2^\omega$, contrary to the assumption.
The same holds for any other oracle of size $<2^{\omega}$.
\end{Proof}

Corollary \ref{cor:subfield}, our natural hope for reproving Yonezawa's result
that $\bH\not\leq_{BSS}\fC$ by the methods of this paper, fails to do so,
for the field generated by $\fC$ does not satisfy the hypothesis there.
It seems counterintuitive that a set of measure $0$ could generate
such a large field, so we prove it here.  (The authors assume that
this fact has been proven long since, and would appreciate a reference for it.)
\begin{lemma}[Folklore]
\label{lemma:generateR}
The Cantor set $\fC$ generates the entire field $\R$.
Indeed, it generates $\R$ as a ring.
\end{lemma}
\begin{Proof}
The argument is best understood by seeing an example.  Here we
begin with an element of $[0,1]$, chosen arbitrarily, in ternary form:
\begin{align*}
&0.2201020001211\ldots\\
=~~&0.2200020000200\ldots\\
+ &0.0001000001011\ldots\\
=~~&0.2200020000200\ldots\\
+ (&0.0002000002022\ldots)\cdot\frac12
\end{align*}
Since $\frac12$ lies in every subfield of $\R$, this shows
that this number is generated from $\fC$ by field operations.
Indeed, since $\frac12=2\cdot\frac14=2\cdot (0.020202\ldots)$,
the number is generated from elements of $\fC$ by ring operations.
The same process can be applied to any element of $[0,1]$,
so $\fC$ generates the entire unit interval, and hence all of $\R$.
(In particular, let $x \in [0,1]$, and write $x$
as $x_1+x_2$, where each non-zero ternary digit of $x_i$ is
equal to $i$.  Now $x'_1 = 2x_1$ is in $\fC$, and
$\frac14$ is in $\fC$.  Thus,
$x = x_2 + (\frac14+\frac14) x'_1$ is in the subring generated by $\fC$.)
\end{Proof}

\section{A Nicer Situation:  the Complex Numbers}
\label{sec:complex}

BSS computation has also been widely considered
on the field $\C$ of complex numbers.  The principal differences
are the algebraic closure of $\C$ and the consequent
impossibility of any order on $\C$ compatible with the field operations.
Of these, the second is probably the more significant difference.
With no order available, BSS machines on $\C$ can only
make comparisons of cell contents under $=$
(and can compute the four field operations, of course).
To help show the importance of this difference, we demonstrate here
how much easier the questions of Section \ref{sec:oracle}
become when considered on $\C$.  The following is the analogue in $\C$
of Theorem \ref{thm:cardinality}.

\begin{thm}
\label{thm:complexcardinality}
If $C\subseteq\C^\infty$ is an oracle set of
infinite cardinality $\kappa <2^\omega$, and
$S\subseteq\R$ is a set with $S\leq_{BSS} C$, 
then either $S$ or its complement must have cardinality $\leq\kappa$.
The same holds for oracles $C$ of infinite co-cardinality
$\kappa<2^\omega$.
\end{thm}
\begin{Proof}
If the machine $M$ with parameters $\zvec$ computes
$\chi_S$ from oracle $C$, consider any two inputs
$x,y\in\C$ which are both transcendental over the field $F$
generated by $\zvec$ and all components of tuples in $C$.
One immediately sees that the computations
of $M^C$ on each of these inputs follow the same path,
with the cell contents at each stage given
by rational functions $f$ with coefficients in $\Q(\zvec)$.
With $x$ transcendental over $F$, the only way for $f(x)$
to lie in $F$ is for $f$ to be constant, in which case $f(y)$
is the same constant.  Thus the oracle questions
in the two computations always yield the same answers,
and so $\chi_S(x)=M^C(x)=M^C(y)=\chi_S(y)$, since the possible
output values for $M^C$ are just $0$ and $1$, which are
both in the field $F$.  So $S$ contains either
all such transcendentals, or else none of them.
Since there are only $\kappa$-many elements
of $\C$ algebraic over the size-$\kappa$ field $F$,
the theorem follows.
\end{Proof}

So, without the order $<$ requiring inputs to be chosen
within $\epsilon$ of other inputs, the result involves
no local bicardinality whatsoever.  We consider this theorem
on $\C$ to be the best starting point for a generalization
to sets $S\subseteq\C^\infty$ BSS-decidable below oracles
of cardinalities $< 2^\omega$.  As stated above, the theorem
is false for such $S$; indeed, in $\C^2$, the zero set of any
finite collection of polynomials in $\C[X,Y]$ is decidable,
and even when the collection contains just
a single nontrivial polynomial,
this set has both cardinality and co-cardinality $2^\omega$.
The correct analogy should be that points in $\C$ should be considered
as varieties there (after all, the singletons are exactly
the irreducible affine varieties in $\C^1$), and that the generalization
of Theorem \ref{thm:complexcardinality} to $\C^n$ should
not concern the cardinality of $S$, but rather the least
possible cardinality of a set of irreducible varieties such
that $S$ is a Boolean combination of those varieties.





\end{document}